# A Triangular Fuzzy based Multicriteria Decision Making Approach for Assessing Security Risks in 5G Networks

**Hisham A. Kholidy**

**kholidh@sunypoly.edu**

**Department of Networks and Computer Security, College of Engineering, State University of New York (SUNY) Polytechnic Institute, Utica, NY.**

***Abstract***

The emerging 5G network is a new global wireless standard after 1G, 2G, 3G, and 4G networks. In comparison to 4G, it has lower latency, larger capacity, and more bandwidth. These network upgrades will have a profound impact on how people throughout the world live and work. The current research investigates mechanisms to protect the 5G networks to meet resilience requirements and to minimize the damage from attacks that do occur. The main contribution of this paper includes: (1) Improving the current 5G security testbed by orchestrating the security services using the OSM and Open stack, Integrating the FlexRan with the testbed components to control and manage the eNodeB, and implementing some real-time security experiments to test and validate the testbed. (2) Develop an intelligent fuzzy method to improve the accuracy of the current Vulnerability Assessment Approach (VAA) using a new approach that integrates the Triangular Fuzzy Numbers (TFNs) and a multi-criteria decision making technique to find the attack graph paths where the attack most probably will propagate in the 5G network.

## 1. INTRODUCTION

5G has the potential to make the world much more connected. It enables a new kind of network that is designed to connect virtually everything and to expand the scope of mobile technology beyond the capabilities of LTE. Devices in locations without access to traditional broadband networks can be deployed using 5G network connectivity. 5G offers higher speeds, lower latency, and increased capacity that make it a potential option for devices where 4G LTE was not a viable option. However, with this growth in systems connected to 5G networks, the 5G network is a potential target for security attacks.

The DoD's 5G strategy implementation plan [1] provides a roadmap for addressing the 5G technology and security aspects. This strategy introduced four recommendations for developing a threat intelligent model that studies the 5G threats and vulnerabilities. Among these recommendations (1) developing techniques to identify, track, and mitigate threats and vulnerabilities that arise from different choices, configurations, and combinations of network equipment, software components, and deployment environments. (2) demonstrating how vulnerabilities in underlying hardware could be exploited by an adversary to impact 5G-enabled capabilities and operations. In the light of these recommendations, a new security method is needed to react to 5G network changes in real-time.

To the best of our knowledge, a few works introduce a real-time security solution that specifically works for 5G networks and considers these networks' real-time scalability and dynamic features due to the lack of publicly available 5G security testbeds, datasets, and attack graphs.

This paper contributes towards:

1) Improving the current 5G security testbed by (a) orchestrating the security services using the OSM and Open stack. We develop a security deployment model that deploys the intrusion detection agents at the edge of the network and connects them to a central management system. The testbed leverages various open-source projects and tools that are used in deploying 5G networks and cybersecurity systems such as OpenStack, Kubernetes, OSSIM, SNORT, and OSSEC. The testbed is flexible to accommodate different aspects of the security analysis from SDN, NFV, intrusion detection to malware analysis. It also includes a programmable traffic generator that can be used to generate real-world traffic load. (b)Integrating the FlexRan with the testbed components to control and manage the eNodeB resources, and (c) run some real-time security experiments to test and validate the testbed. The improved security testbed enables a multi-tenancy for the deployment of virtualized end-to-end network slices. It also allows researchers to conduct various security analyses, compare the accuracy of different machine learning approaches and defense mechanisms, and execute a simulated cyberattack.
2) Improve the accuracy of our original VAA [2]. The original VAA uses the TOPSIS (Technique for Order of Preference by Similarity to Ideal Solution) [3, 4] as a multi-criteria decision making technique to find the attack graph path with the lowest attacker costs. This path indicates where the attack most probably will propagate. The improved approach (FuzVAA) integrates the original VAA with the Triangular Fuzzy Numbers (TFNs). The reason for integrating the TOPSIS with the TFNs is that the classical TOPSIS alone is found to have some vagueness and is not sufficient to arrive at a solution because of its higher dimensionality particularly when we used it with a large scale 5G network with a large number of UEs. The accuracy, scalability, and performance of the FuzVAA technique are tested and evaluated using the improved 5G security testbed.

## 2. DISCUSSION OF PROBLEM

The extensive use of machine learning and game-theoretic approaches in the military sector has changed the face of the battlefield and the nature of war. To the best of our knowledge, none of the current work introduces a real-time holistic framework that specifically works for 5G core networks and integrates the vulnerability analysis, risk assessment, and mitigation in the 5G systems. This is due to the lack of publicly available real-time 5G testbeds, datasets, and attack graphs.

Most of the current state of the art [15-20] focus on either the SDN or NFV security and do not consider the 5G core challenges such as (1) performance monitoring, (2) scalability, (3) orchestration and management, and (4) heterogeneous network support and integration of the SDN, NFV, and edge computing. Few works study the vulnerability analysis and risk assessment in 5G Networks. However, they are still at an early stage. Current traditional risk assessment and vulnerability analysis approaches that are based on ISO/IEC 2700 series [21] and NIST-SP800 [22] do not take into account the 5G core design principles. In [23], the authors introduced an intrusion prevention system that employed five layers of 5G systems to detect the flow table overloading attack. However, this work is more specific to a particular attack category and does not consider the rest of the 5G attack vectors. Furthermore, it lacks the vulnerability analysis of the 5G core components. In [24-38, 48, 49], we introduced a security framework that detects attacks in cloud-based systems. This framework integrates behavior-based and signature-based systems and uses machine learning and artificial intelligence techniques to develop self-defensive techniques for cloud systems.

Based on this survey, none of the current approaches considers the evaluation of the security risk for the underlying resources that the slices use nor offers a dynamic allocation scheme to re-allocate previously allocated slices to make more efficient use of available resources (or newly added resources).

In [2] we developed the VAA that dynamically assesses and analyzes the vulnerabilities in the 5G networks using the TOPSIS. Experiments that were recently conducted on large scale 5G testbed proved that the TOPSIS has some vagueness and is not sufficient to arrive at a solution because of its higher dimensionality that negatively impacts its accuracy and performance.

The use of ranking fuzzy numbers in soft computing plays an important role in converting the classical decision-making problem into a fuzzy decision-making one; therefore, improving the final efficiency and performance of the ranking of the decision alternatives. Section 3.2 discusses the new FuzVAA that integrates the ranking fuzzy numbers and the classical TOPSIS to address the vagueness issues of the classical TOPSIS.

An efficient security framework that works for the 5G core networks should address the following top security challenges that service providers need to tackle to make 5G services a successful business [44, 45]:

1) **Network slicing and virtualization bring new risks.** Service providers need to consider how well the virtualization layers and network slices are isolated from each other.
2) **Heterogeneity and complexity of a variant infrastructure** that the 5G system includes.
3) **Automated threat detection and mitigation.** For previous generations, manual interventions for mitigating threats may be fine. But with 5G now and the fact that the threat landscape continues to grow in volume, velocity, and sophistication particularly for the 5G core network, manual operations are not enough.
4) **The coexistence of 4G and 5G requires scalable and higher performance security.** 5G will evolve side by side with 4G. The increase in bandwidth from 4G eNodeB to 5G gNodeB will cause significant increases in performance and scale requirements that the current security infrastructure may not be able to handle.
5) **Distributed edge clouds open up new attack surfaces.** In the 5G architecture, IP connectivity will terminate at the edge of the operator if proper security mechanisms such as encryption and firewalls are not in place.
6) **Threats to the availability and integrity of networks** will become major security concerns: in addition to confidentiality and privacy threats, with 5G networks expected to become the backbone of many critical IT applications, the integrity, and availability of those networks will become major national security concerns.

## 3. METHODOLOGY

### 3.1 The Security Testbed

In this section and following subsections, we highlight the security testbed, explain how the network slicing is implemented in the testbed using the Network Function Virtualization (NFV) features, discuss our security deployment model and the orchestrated security services, introduce the threat model and attack scenarios that are used in the experiments, and discuss a cybersecurity use case to validate the Testbed.

Our testbed leverages various open-source projects and tools that are used in deploying 5G networks and cybersecurity systems such as OpenStack [5], Open-Source MANO (OSM) [6], Open5GS [7], FlexRAN (Mosaic5G) [8], OSSIM [9], SNORT [10], and OSSEC [11]. The improved security testbed accommodates different aspects of the security analysis from SDN, NFV, intrusion detection to malware analysis, and the Radio Access Network (RAN). The 5G security testbed allows researchers to (1) conduct various security analyses, (2) build security datasets and 5G based attack graphs, (3) benchmark and compare the accuracy and performance of different machine learning approaches and defense mechanisms, and (4) execute simulated cyberattacks. The testbed aims to provide the following capabilities:

(1) a programmable traffic generator that can be used to generate real-world traffic load.
(2) multi-vendor implementation by abstracting the underlying platforms and enabling the configuration of different vendor-specific platforms.
(3) monitoring and management of both the access and core network resources.

Fig. 1 shows a high-level overview of our 5G network security testbed where SDN and NFV are the foundation of the architecture. Our testbed uses OpenStack, Open-Source MANO (OSM), and Open5GS to deploy 5G core services and the service-based architecture. It also includes FlexRAN for the SDN-RAN controller and OAI eNB for achieving end-to-end network slicing. We briefly list the components of the 5G core testbed as depicted in Fig. 1.:

- **OpenStack** is an open-source hypervisor platform that uses pooled physical and virtual resources to deliver Infrastructure-as-a-service (IaaS). It supports cloud computing and OpenFlow to manage the orchestration of virtual and physical network devices, support the virtualization and control layers in the middle.

- **The Open-Source Network Function Virtualization Management and Orchestration (OSM)** handles the management and orchestration of NFV layers. OSM enables the creation of network services with programmatic ease. It has two principal elements for building a network service: (1) VNF packages, which represent virtual network functions, and (2) NS packages, representing network services. Once a network service is modeled and defined, OSM can spin up and deploy an instance to one or more hypervisor platforms, such as OpenStack in our case, see Fig. 2.

- **Open5GS** integrates with 5G New Radio Stand-Alone (SA) base stations and user equipment enabling immediate demonstration of different features and applications and supporting the current need to have a flexible 5G Core Network in addition to the evolved EPC one (used for LTE and 5G NSA).

- **FlexRan** platform is made up of two main components: the FlexRAN Control Plane and FlexRAN Agent API. The FlexRAN, see Fig. 1. control plane follows a hierarchical design and is in turn composed of a Master Controller that is connected to several FlexRAN Agents, one for each base station (eNodeB in 4G LTE). The control and data plane separation are provided by the FlexRAN Agent API which acts as the southbound API with FlexRAN control plane on one side and eNodeB data plane on the other side. The FlexRAN protocol facilitates the communication between the master controller and the agents.

- **Testbed Visibility**

  Network visibility is a granular awareness of all traffic and user activities on the network, driving observability and helping identify possible bad actors and malicious activity. Our testbed employs common observability strategies in the monitoring of distributed systems. Structured logging of packet and data flows, servers, services, and user equipment are all carried out with open-source monitoring and intrusion detection technology such as the Snort application, a Mobile Agent Intrusion Detection System (MA-IDS), Open source HIDS Security (OSSEC), and Open-Source Security Information Management (OSSIM).

- **Programmable Traffic Generators**

Our user equipment portion of the testbed contains a virtual cluster of mobile device images that run the Android system (BlueStack and Android Studio). This cluster is capable of spinning up virtual mobile devices to produce real-time load on the system. The virtualization hosts and guests run on a base image of Ubuntu Server 20.04 LTS and Windows 10 enabling users to programmatically script both attack and response scenarios in almost any Linux and Windows compatible programming language.

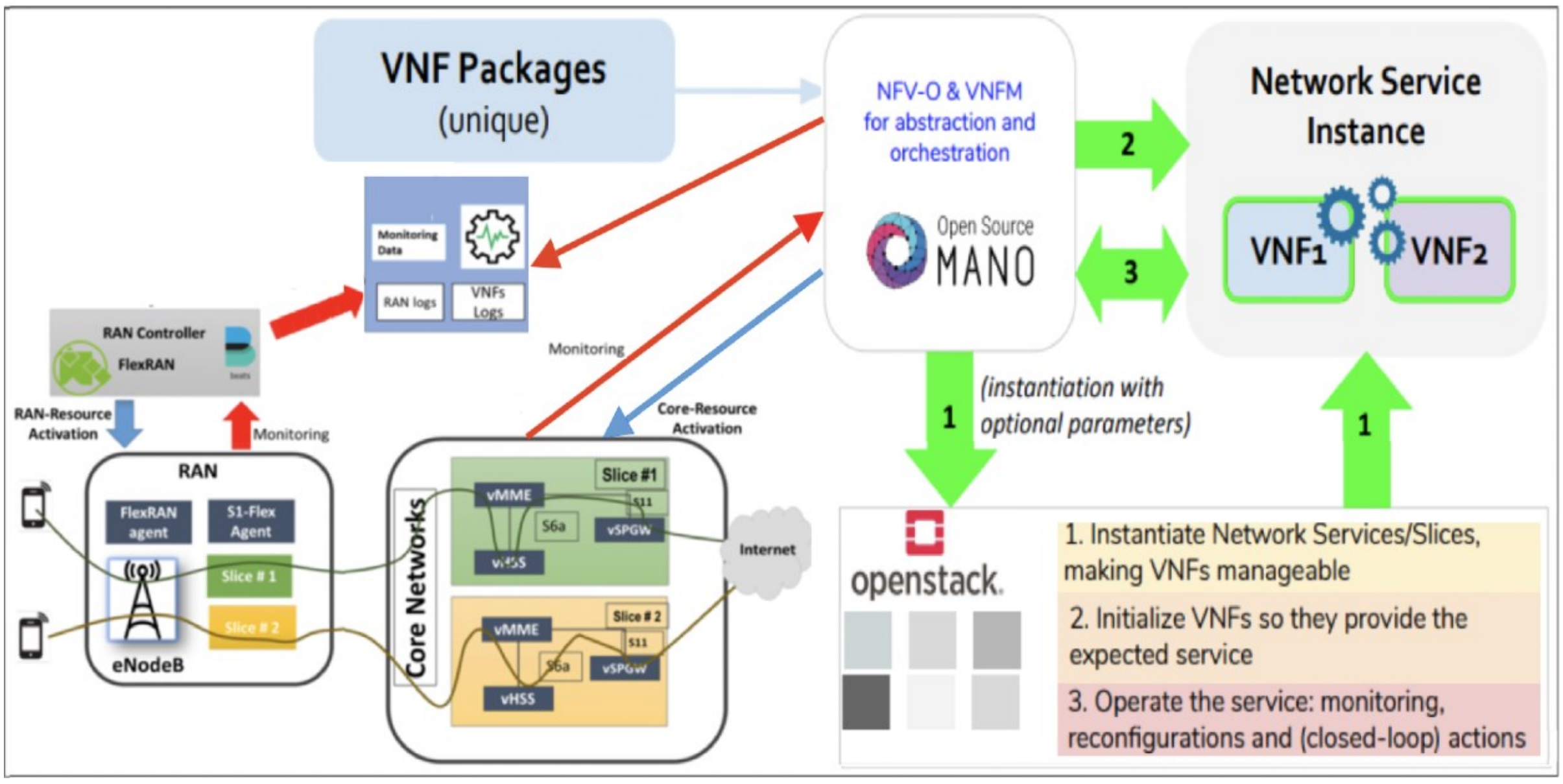

Fig. 1: A high-level Overview of Our Security Testbed Architecture

Table 1: Test-bed component configurations and parameters

| Component | System Parameters |
|---|---|
| **OSM, OpenStack, and Open5GS** | OS: UBUNTU 20.04 LTS RAM: 128 GB CPU: 32 Cores 2.10 GHz<br>SSD: 3 TB (RAID 5) OSM Version: 9.1. MicroK8s Version: 1.19<br>OpenStack Version: Wallaby. Open5GS Version: 2.3.<br>OAI-CN Version: 1.0 |
| **FlexRAN** | OS: UBUNTU 20.04 LTS. RAM: 32 GB. CPU: 4 Cores 2.33 GHz<br>SSD: 2 TB (RAID 5) |
| **SDR USRP B210** | Frequency Range: 70 MHz-6 GHz Channels: 2TX * 2RX |

### 3.1.1 Network Slicing and Network Function Virtualization

To understand the proposed security deployment model, we highlight two main concepts, the Network Function Virtualization, and network slicing.

The NFV MANO covers the orchestration and lifecycle management of physical and/or software resources that support the infrastructure virtualization, and the lifecycle management of VNFs. NFV Management and Orchestration focus on all virtualization-specific management tasks necessary in the NFV framework. Network Function Virtualization leverages high-performance computing and enterprise virtualized environments to divorce network functions from the substrate hardware and implement those functions via software that runs on commercial off-the-shelf high-volume servers, storage, and switches. These are known as Virtual Network Functions (VNF). The architecture underpinning NFV MANO, shown in Figure 2, relies on three principal elements [13]:

1- Virtualized Network Functions (VNF) - A VNF is a programmatic emulation of a formerly physical network function designed for instantiation on Network Function Virtualization Infrastructure.

2- Network Function Virtualization Infrastructure (NFVI) - The NFVI is the complete infrastructure comprised of all the hardware and software that VNFs deploy on.

3- NFV Management and Orchestrator (NFV-MANO) - NFV-MANO is functionally broken down further into that of the NFV Orchestrator (NFVO), the VNF Manager (VNFM), and the Virtualized Infrastructure Manager (VIM).

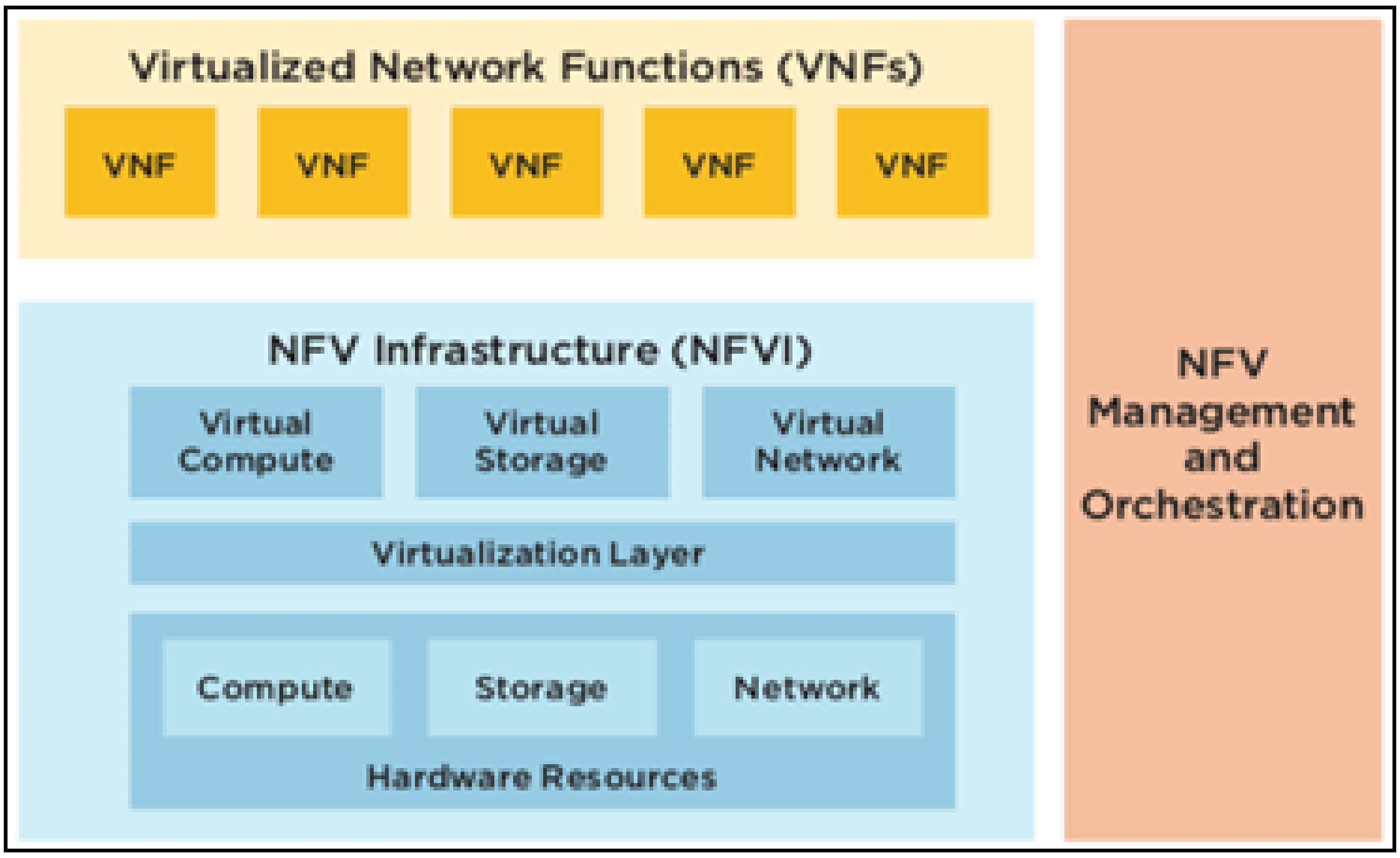

Fig. 2: NFV MANO Architecture

### 3.1.2 The Security Deployment Model

The proposed testbed uses OSM and OpenStack to deploy 5G service-based architecture. This different subset of the system can be focused to perform targeted security analysis. Fig.3 depicts our security deployment model. From the top, the network slices that service enhanced Mobile Broadband (eMBB) phones, tablets, and other user equipment (UE) will contain OSSEC Agents deployed as NVF / NS, and the UE will have MA-IDS installed. The massive Machine Type Communications (mMTC) that services IoT devices, which are less likely to have security or agent software, will also have OSSEC Agent deployed to its slices.

All OSSEC agents talk back through the radios using FlexRAN to the OSSEC server to analyze and make decisions about the severity of OSSEC alerts. The alerts fired by the OSSEC server, MA-IDS Agents, and any alerts from the adjacent SNORT that is sniffing the bidirectional traffic as it is passed to and from slices, and any risk assessments from OSSIM agents are forwarded to the OSSIM server at the CORE to normalize, correlate, and analyze them to compute the overall risk in the 5G testbed. OSSIM server is dedicated to processing security events and providing command and control back up the stack from CORE, to COMPUTE, to EDGE, and through UE, using OSM to orchestrate dynamic changes. At present, the stack is contained in the bottom system ("model3") in our testbed and it functions as CORE + COMPUTE + EDGE.

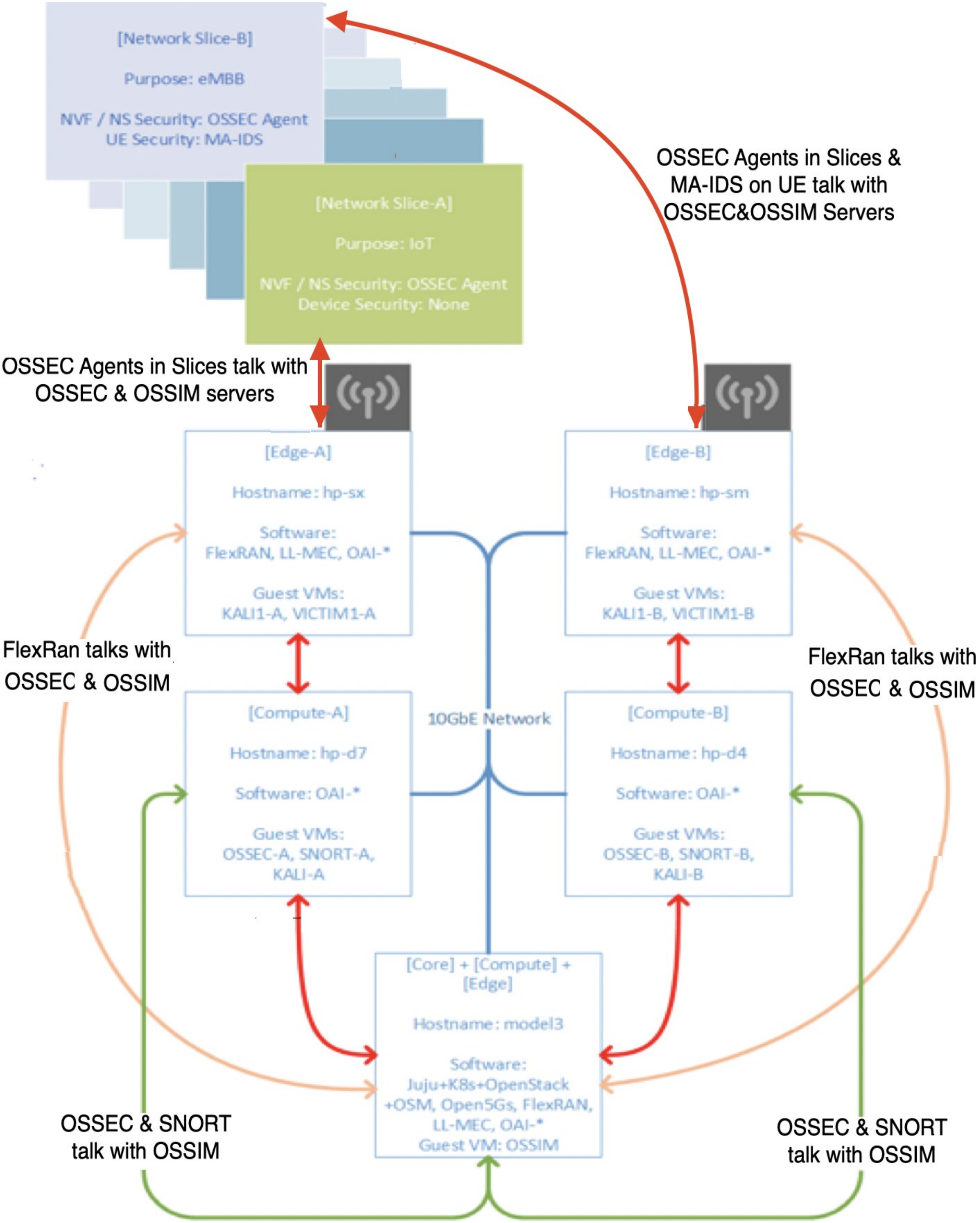

Fig. 3: The Security Deployment Model

### 3.1.3 Threat Modeling and Attack Scenarios

SDN and NFV bring a lot of benefits that are crucial in realizing the 5G network but they also bring a lot of security risks. Some of the risks might include vulnerable software on VNF, DoS attack on MANO, hypervisor hijacking, exploiting SDN controller, etc. There are many variants of threats to 5G infrastructure, but they can be organized according to impact into three, broad categories:

- Denial of Service: Attacks that cause service outages through either virtual or physical means. Examples include hardware vandalism, radio signal jamming, and traffic flooding.
- Access Breach: Attacks that lead to unauthorized system access, manipulation, or a data breach. Examples include malicious changes to system configurations, exploiting software vulnerabilities, communications hijacking, and disclosure of sensitive data.
- Integrity Compromise: Attacks that impact infrastructure integrity through hardware, software, communications, or operational tampering. Examples include implanting malicious hardware components in system devices, altering operating system code, modifying data, and circumventing organizational security policies.

Besides the abovementioned threats, there are three additional attack schemes (I, S, P) enabled by the integration of Mobile Edge Computing (MEC) and 5G networks that we focus on in our study, see Fig. 4, namely:

(1) **I**nsecure mobile backhaul network. Data exchanged between MEC nodes often traverse insecure shared backhaul that is vulnerable to MITM attacks, including eavesdropping and spoofing. Such attacks can also come from edge nodes connected to the public internet through the edge Firewall Interfaces (e.g., SGi/N6).

(2) **S**hared infrastructure with third-party applications. MEC nodes can be opened to allow authorized participants to deploy applications/services to other users. However, poorly designed applications can create opportunities for attackers to invade the system and pose threats to the network applications running on the platform.

(3) **P**rivacy Leakage Illegitimate access to the Multi-access MEC system. In this case, an attacker can compromise the service infrastructure and the network hampering information privacy. An attacker's ability to access the information stored at the edge system's upper layers poses a serious concern for privacy leakage.

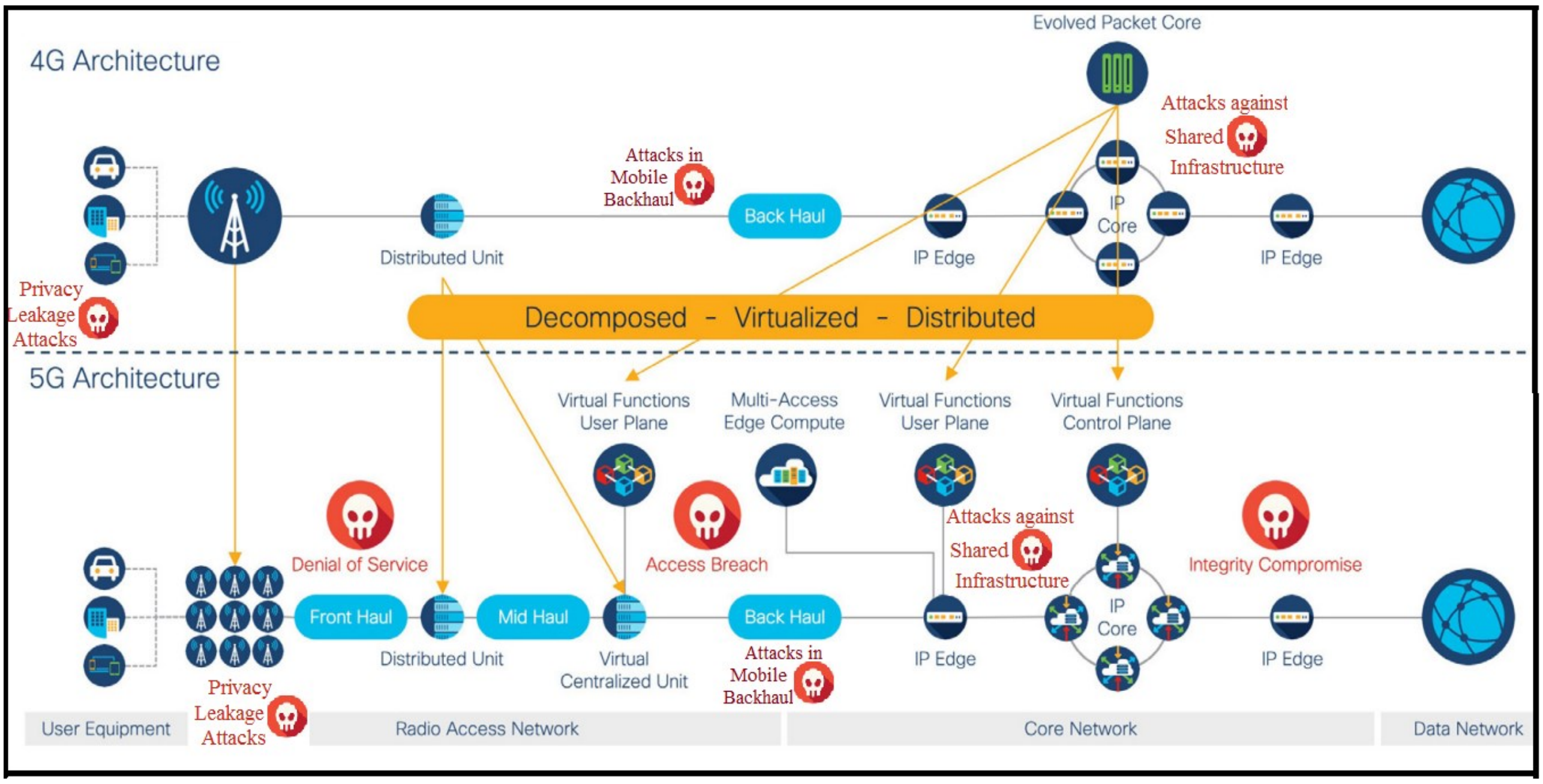

Fig. 4: Attacks in the 5G Network

### 3.2 The FuzVAA

In the following three subsections, we introduce some preliminaries about the triangular fuzzy numbers then we highlight the previously developed VAA, finally we introduce the new FuzVAA approach.

#### 3.2.1 Preliminaries

In this section, we will cover some basic definitions and preliminaries about the triangular fuzzy numbers that are used by the FuzVAA.

**Definition 1**: A fuzzy number is a generalization of a regular, real number in the sense that it does not refer to one single value but rather to a connected set of possible values, where each possible value has its own weight between 0 and 1. This weight is called the membership function ($\mu_A$).
$\mu_A$: $X \rightarrow [0,1]$. This assigns a real number $\mu_A(x)$ in the interval $[0, 1]$, to each element $x \in A$, where the value of $\mu_A(x)$ at $x$ shows the grade of membership of $x$ in $A$.

**Definition 2:** A fuzzy number is a convex normalized fuzzy set of the real line $R$ whose membership function is piecewise continuous.

**Definition 3**: Triangular fuzzy number [47] A can be defined as a triplet (a, b, c). Its membership function is defined as, see Fig. 5:

$$\mu_{\tilde{A}_i}(x) = \begin{cases} \frac{(x - a)}{(b - a)} & a \leq x \leq b \\ \frac{(c - x)}{(c - b)} & b \leq x \leq c \\ 0 & \text{other wise} \end{cases}$$

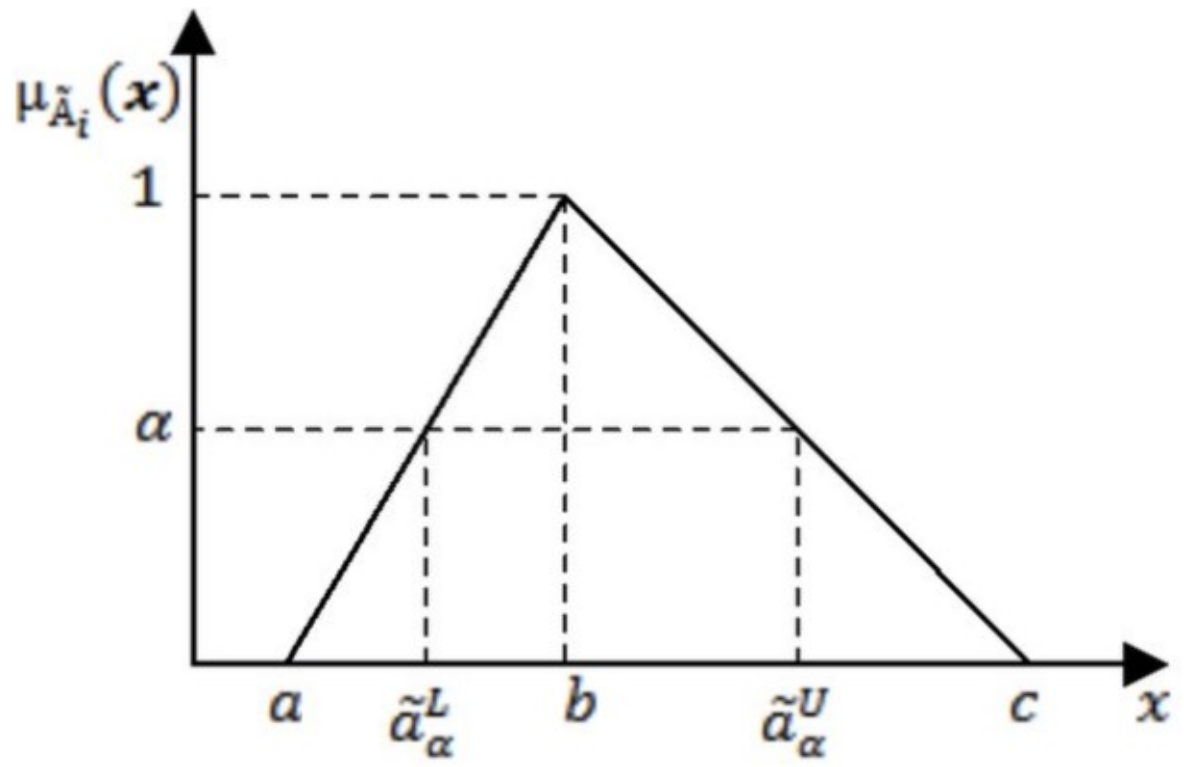

Fig. 5**:** Triangular fuzzy number membership function

Where: $a, b, c$ are real numbers and $a < b < c$. The value of $x$ at b gives the maximal grade of $\mu_{\tilde{a}}(x)$, i.e., $\mu_{\tilde{a}}(x)= 1$; it is the most probable value of the evaluation data. The value of $x$ at $a$ gives the maximal grade of $\mu_{\tilde{a}}(x)$, i. e., $\mu_{\tilde{a}}(x) = 0$; it is the least probable value of the evaluation data.

**Definition 4**: The distance between two fuzzy triangular numbers $\tilde{a} = (a, b, c)$ and $\tilde{b} = (a', b', c')$ is given using the vertex method by: $dist(\tilde{a}, \tilde{b}) = \sqrt{\frac{1}{3}[(a-a')^2 + (b-b')^2 + (c-c')^2]}$

**Definition 5**: The multiplication of two fuzzy triangular numbers $\tilde{a} = (a, b, c)$ and $\tilde{b} = (a', b', c')$ is given using the vertex method by: $(a, b, c)\ (.)\ (a', b', c') = (a \times a', b \times b', c \times c')$.

### 3.2.2 The Original VAA.

Our original VAA [2] develops (1) a scalable attack Graph Generator (GG) model. (2) A new dynamic Vulnerability Analysis (VA) model that hierarchically analyzes the generated attack graphs using the TOPSIS [3, 4] to model the multiple criteria decision-making problem in the 5G core dynamic environment. TOPSIS is based on the concept that the chosen alternative should have the shortest geometric distance from the positive ideal solution and the longest geometric distance from the negative ideal solution. Ideal solutions in the current context refer to the lowest attacker cost of actions that denotes the lowest attacker efforts to exploit a certain vulnerability. E.g., in Fig. 6, if the computed TOPSIS cost of exploitation of CVE2004-0417 is lower than CVE2002-0392 and CVE2004-0415, this means if the attacker started exploiting CVE2004-0417 rather than the other vulnerabilities this will be considered a positive ideal solution. The following steps summarize the VA approach.

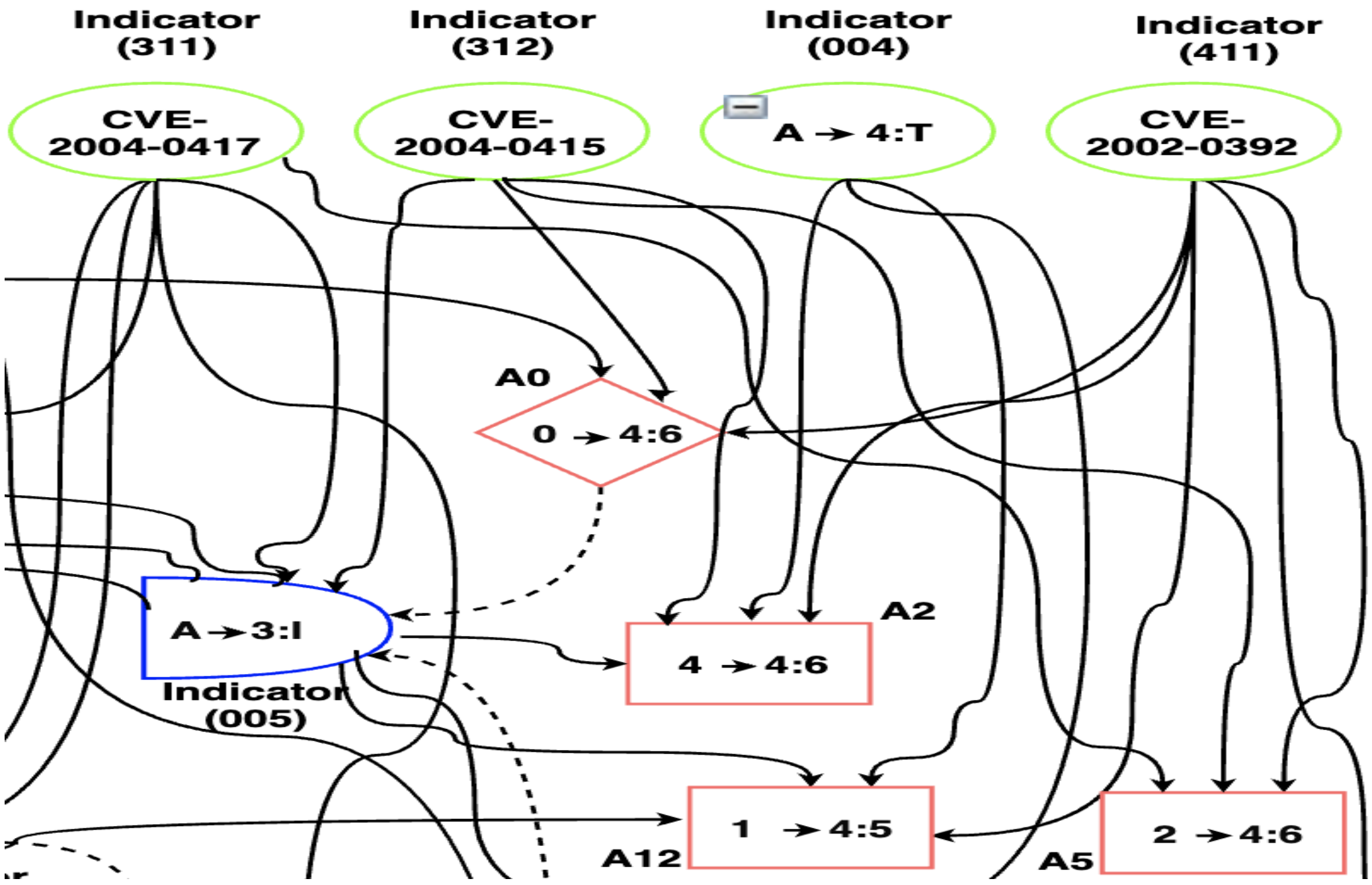

Fig. 6: Part of an example of the generated attack Graph

**Step 1: Develop a scalable attack Graph Generator (GG) model.** This model is based on the security attack vector that focuses on the attacks and threats that may harvest intelligence from the 5G network resources, states, and flows as a result of the integration of the NFV and SDN. The basic idea underlying this model is that the attacker action cost is under the constraint of certain vulnerability and network dynamic factors/indicators of the 5G network such as latency, accessibility, and other factors described in [39]. The vulnerability factors refer to the Common Vulnerability Scoring System (CVSS) factors/indicators namely Base, Temporal, and Environmental. Each of these factors is a composite of other several factors described in [40]. We model this problem as a multi-objective decision-making problem as follows.

**(1) Create the GG three-layer hierarchical model** based on the vulnerability and dynamic network factors, see Fig. 7. The attack graph is modeled based on these factors. An attack graph is defined as a tuple $G = (A, S, T)$, where $A$ is a set of attack actions, $S$ is a set of system states, T is a set of targets that the attacker tries to achieve. An attack graph GG consists of a set of nodes of four types, see Fig. 7: (1) attack-step nodes (circular-shaped AND-nodes). Each node in this set represents a single attack step that can be carried out when all the predecessors (preconditions to the attack which are either configuration settings or network privileges) are satisfied. (2) Privilege nodes (diamond-shaped nodes). Each node in this set represents a single network privilege. The privilege can be achieved through any one of its predecessor AND node which represents an attack step leading to the privilege. Each node in this set represents a fact about the current network configuration that contributes to one or more attack possibilities (subaction). (3) Configuration nodes (circular-shaped). Each node in this set represents an initial vulnerability, configuration settings, or network privileges that are known to be true and have no variance in probability. (4) Final step nodes (rectangular-shaped). Each node in this set represents a final exploit action against a certain vulnerability.

**(2) Construct a pair-wise evaluation matrix *M*,** see Fig. 8, based on the attack graph. After that, we compute the combinatorial weights ($W^i$) which refer to the weight of the impact of each layer's dynamic factors, in the GG three-layer model, on the attacker decision as given in Eq.1.

$$W^i = (W_j^{iL})_{j=1\rightarrow n} W^L \qquad (1)$$

Where $i$ is the GG hierarchical layer index $\in \{1,2,3\}$, $j$ refers to the dynamic factors, and $W^L$ is the criteria layer combinational weight vector which is computed as given in Eq.2.

$$W^L = M*W \qquad (2)$$

*Where,* $W$ is the relevant normalized characteristic vector/eigenvector $= \lambda_{max} * W$, *for all* $w = (w_1, w_2, w_3, \ldots. w_n)$. $\lambda_{max}$ *is the largest* eigenvalue of matrix $M$.

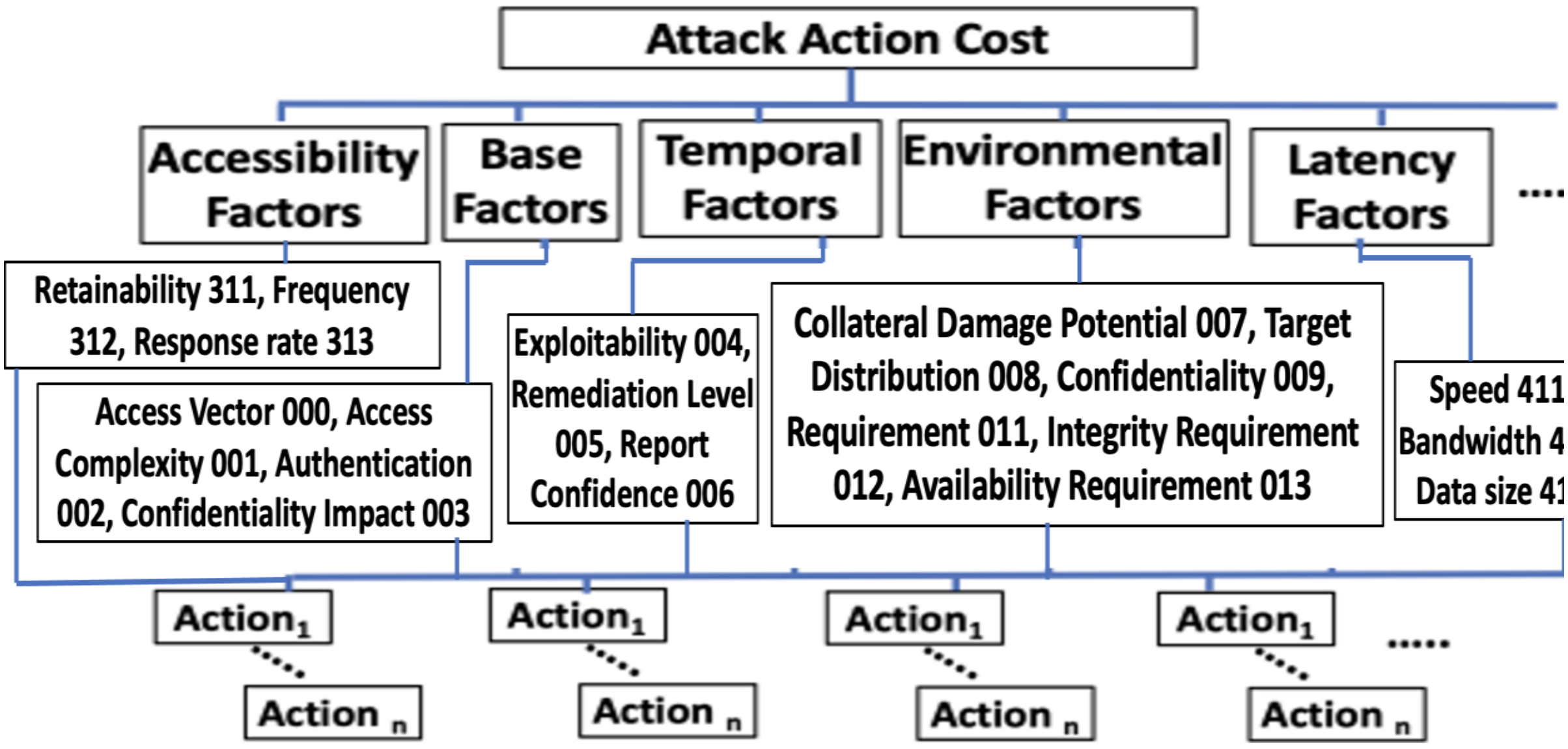

Fig. 7: The Hierarchical GG with corresponding factors' codes

$$(M)_{n\times n} = \begin{bmatrix} 1 & a_{12}\cdots & a_{1n} \\ a_{21} & 1 \;\cdots & a_{2n} \\ \cdots & \cdots \quad \cdots & \cdots \\ a_{n1} & a_{n2}\;\cdots & 1 \end{bmatrix}$$

Fig. 8: The *M* pair-wise Matrix

**Step2: Develop the new dynamic Vulnerability Analysis (VA) model.** To compute the attack cost of actions, we will apply the TOPSIS approach as follows.

**(1) Normalize the pair-wise decision matrix *M*** to form the normalized decision matrix *N* as given in Eq.3.

$$N_{ij} = (N_{ij})_{m\times n} \qquad (3)$$

Where, $N_{ij} = \frac{M_{ij}}{\sqrt{\sum_{j=1}^{n} M_{ij}^2}}$, $i = 1,2,\ldots.,m, j = 1,2,\ldots.n$

**(2) Calculate the weighted normalized decision matrix and the best and worth alternatives.**
The weighted normalized decision matrix $D = N\times W$. *The best alternative* $E^+$ *and the worst alternative* $E^-$ *are defines in* Eq. 4 and 5 respectively**.**

$$E^+ = (E^+_1, E^+_2, E^+_3, \ldots\ldots, E^+_n) \tag{4}$$

$$E^- = (E^-_1, E^-_2, E^-_3, \ldots\ldots, E^-_n) \tag{5}$$

Let's define the benefit criteria as B and the cost criteria as C. The value of E+ and E- can be calculated using Eq. 6 and 7 respectively.

$$e_i^+ = [\,MAX\,(d_{b,c}) \mid I \in B],\ [\,MIN(d_{b,c}) \mid i \in C\,] \tag{6}$$

$$e_i^- = [\,MIN\,(d_{b,c}) \mid i \in B],\ [\,MAX(d_{b,c}) \mid I \in C\,] \tag{7}$$

**(3) Calculate the cost of the attacker's actions.** We use the L2-distance defined by the TOPSIS approach to calculate $L2_i^+$, the distance between the target alternative *i* and the best condition $E^+$ as given in Eq. 8, and $L2_i^-$, the distance between the target alternative *i* and the worst condition $E^-$ as given in Eq. 9.

$$L2_i^+ = \sqrt{\sum_{k=1}^{n} (e_{i,k} - e_k^+)^2} \tag{8}$$

$$L2_i^- = \sqrt{\sum_{k=1}^{n} (e_{i,k} - e_k^-)^2} \tag{9}$$

Based on the $L2_i^+$ and $L2_i^-$ distances, we compute the similarity to the worst condition as the cost of the attacker's actions ($Atc_{Cost}$) as shown in Eq.10.

$$Atc_{Cost}(i) = \frac{L2_i^+}{L2_i^+ + L2_i^-} \tag{10}$$

Where $i \in \{1,2\ldots,m\}$ is the actions the attacker can choose from *m* possible actions. Using the hierarchical GG in Fig. 7, we give a simple demonstration for the decision matrix of attacker actions compared to the network indicators (the network components where the attacker may start its exploitation), see Table. 1. The full case study of this example is detailed in [2]. The computed attack graphs, actions, and the costs of these actions can be used by an intrusion response system to model the security reciprocal interaction between it and the attacker and can help in deploying the best countermeasures to mitigate the attacks in the 5G core networks.

Table 1: Attacker Decision Matrix

| Attacker Goal | Exploitation Starting Point | | | |
|---|---|---|---|---|
| | | CVE-2004-0417 | CVE-2004-0415 | CVE-2002-0392 |
| | ***I:*** disruption for NFVI Services | A5 | A5 | A5 |
| | ***S:*** illegitimate access to Shared SDN | A12 | A0-A12, A12, | A12 |
| | ***P:*** illegitimate access to the RAN | A2 | A0-A2, A2 | A2 |

### 3.2.3 The FuzVAA: the Hybrid TFN and TOPSIS Approach

The new proposed hybrid approach integrates the TOPSIS with the TFN. This approach uses the same attack Graph Generator (GG) and the three-layer hierarchical model that is based on the vulnerability and dynamic network factors described in Section 3.2.2. In the following, we describe the steps of the proposed approach.

**Step1: Construct the fuzzy decision matrix *M* and Compute the Aggregated Fuzzy Ratings.**
Let's say the decision pool has *N* members. If the fuzzy rating and importance weight of the $N^{th}$ decision maker (attacker decision) about the $i^{th}$ alternative on the $j^{th}$ criterion (the weight of the criterion refers to the weight of the impact of each layer's dynamic factors, in the GG three-layer model in Fig. 7, on the attacker action/decision) are $x\tilde{}_{ij}^{N} = \left( a\tilde{}_{ij}^{N}, b\tilde{}_{ij}^{N}, c\tilde{}_{ij}^{N}, \right)$and $w\tilde{}_{j}^{N} = \left( a'^{N}_{j}, b'^{N}_{j}, c'^{N}_{j} \right)$ respectively, where *i= 1, 2, 3, ……………,m* and *j= 1, 2, 3,…………..,n* then aggregated Fuzzy ratings $x\tilde{}_{ij}$ of alternative *i* with respect to each criteria *j* are given by $x\tilde{}_{ij}= (a_{ij}, b_{ij}, c_{ij})$ where:

$$a_{ij} = {}^{Min}_{N}\{a_{ij}^{N}\}, \; b_{ij} = \frac{1}{N}\sum_{N=1}^{N} b_{ij}^{N}, \; c_{ij} = {}^{Max}_{N}\{c_{ij}^{N}\} \quad (11)$$

In the same way, the aggregated Fuzzy weights of each criterion are calculated as $w\tilde{}_{j}= (a'_{j}, b'_{j}, c'_{j})$ such that:

$$a'_{j}= {}^{Min}_{N}\{a'^{N}_{j}\}, \; b'_{j}= \frac{1}{N}\sum_{N=1}^{N} b'^{N}_{j}, \; c'_{j} = {}^{Max}_{N}\{c'^{N}_{j}\} \quad (12)$$

The fuzzy decision matrix has each entry of the TFN as given below:

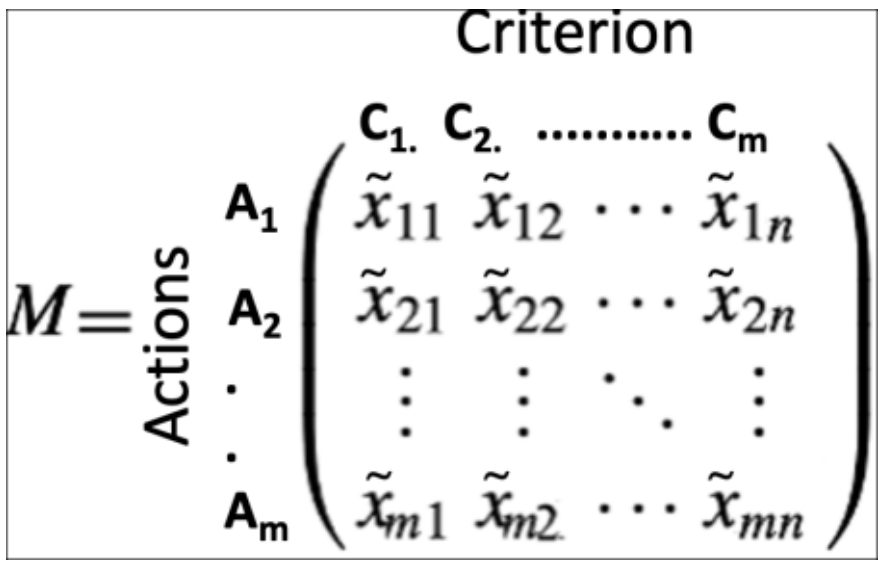

$$W\tilde{} = (w\tilde{}_{1}, w\tilde{}_{2}, w\tilde{}_{3}, w\tilde{}_{4}, \ldots\ldots\ldots w\tilde{}_{n}) \quad (13)$$

Where $X\tilde{}_{ij} = (x\tilde{}_{1,1}, x\tilde{}_{1,2}, \ldots . x\tilde{}_{1,n}, x\tilde{}_{2,1}, x\tilde{}_{m,1}, x\tilde{}_{m,2} \ldots\ldots, x_{mn})$, *i= 1, 2, 3,..., m; j = 1, 2, 3,..., n*, represents the number of alternatives attack actions and decision criteria, respectively and $X\tilde{}$and $W\tilde{}$are triangular Fuzzy numbers representing linguistic variables.

**Step2: Construct the normalized decision matrix** $\bar{M}$ using *M* entries. Unlike the classical TOPSIS method described in Section 3.2.2, which uses the $\lambda_{max}$(the largest eigenvalue of matrix *M*) to compute the weight of the criterion, we use the linear scale transformation to simplify the transformation of various criteria scales into a comparable scale. Normalized fuzzy weights make it possible to model mathematically an uncertain division of a unit into *n* fractions. The normalized fuzzy weights preserve the property that the ranges of normalized triangular fuzzy numbers belong to (0, 1). Fig. 9 illustrates an example of normalized fuzzy weights that uses a linguistic scale that is depicted in Table 2. Each criterion and alternative are evaluated by the triangular numbers according to this linguistic scale.

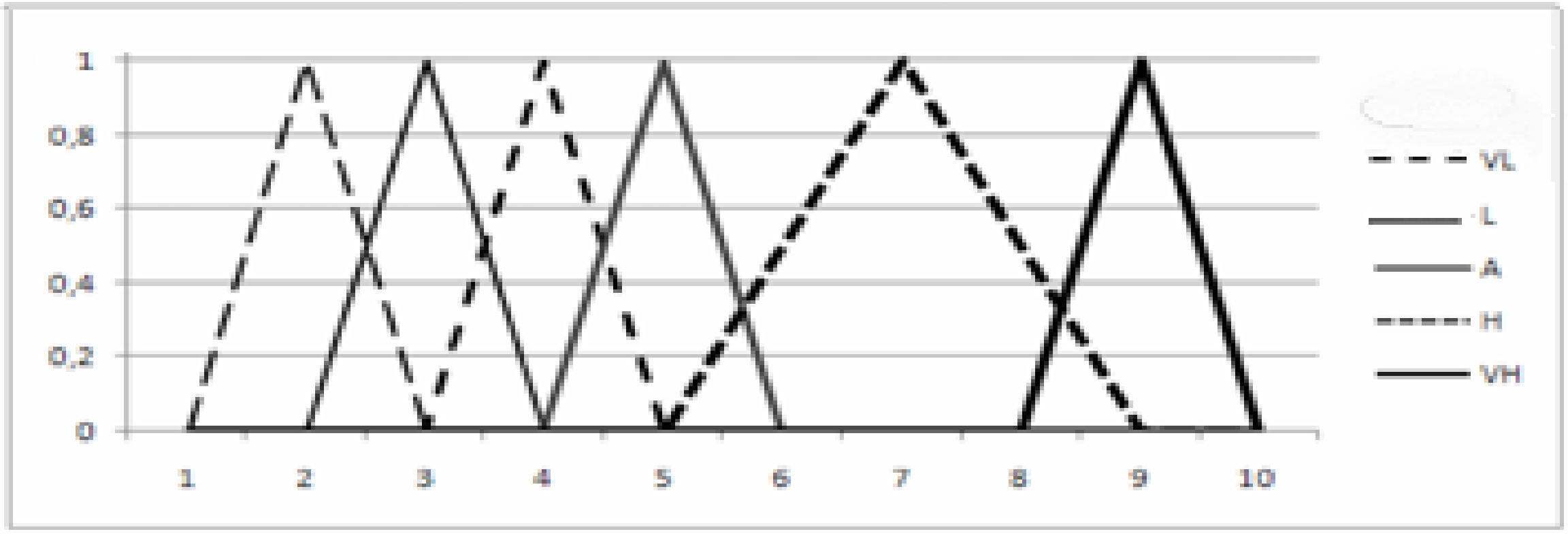

Fig. 9: An example of normalized fuzzy weights using a linguistic scale.

Table 2: Linguistic Variables for the Criterion Weight Impact and the Rating

| Linguistic Variable | Fuzzy Value |
|---|---|
| Very Low (VL) | 1,1,3 |
| Low (L) | 1,3,5 |
| Average (AV) | 3,5.7 |
| High (H) | 5,7,9 |
| Very High (VH) | 7,9,9 |

Thus, the normalized fuzzy decision matrix ($F^{\sim}$) is given as:

$$F^{\sim} = [f^{\sim}_{ij}]_{m*n},\ i=1,2,3,\ldots\ldots,m;\ j=1,2,3,\ldots\ldots,n \tag{14}$$

Where,

$$f^{\sim}_{ij} = (a_{ij}/\ c_j^{Max},\ b_{ij}/\ c_j^{Max},\ c_{ij}/c_j^{Max}) \text{ and, } c_j^{Max} = \text{Max } c_{ij} \quad \textbf{(Benefit Criteria)} \tag{15}$$

$$f^{\sim}_{ij} = (a_j^{Min}/c_{ij},\ b_j^{Min}/c_{ij},\ c_j^{Min}/c_{ij}) \text{ and, } a_j^{Min} = \text{Min}(a_{ij}^{Min}),\ b_j^{Min} = \text{Min}(b_{ij}^{Min}),\ c_j^{Min} = \text{Min}(c_{ij}^{Min}) \quad \textbf{(Cost Criteria)} \tag{16}$$

The benefit criteria from the attacker perspective include high exploitability, high latency, low speed… etc. The cost criteria include long exploit time, low latency, low speed… etc.

The weighted normalized fuzzy decision matrix $M^{w\sim}$ is computed by multiplying the weights ($w_j^{\sim}$) of evaluation criteria with the normalized fuzzy decision matrix $F^{\sim}_{ij}$ as given in Eq. 16.

$$M^{w\sim} = [M^{w\sim}_{ij}]_{m*n}, \text{ where: } i = 1, 2, \ldots.., m;\ j = 1, 2, 3, \ldots.., n;\ M^{w\sim}_{ij} = f^{\sim}_{ij}\ (.)\ w_j^{\sim} = (a''_{ij}, b''_{ij}, c''_{ij}) \tag{17}$$

**Step 3: Calculate the fuzzy positive ideal alternative $E^{\sim+}$ and the fuzzy negative ideal alternative $E^{\sim}$** as shown in Eq. 18 and 19 respectively**.**

$$E^{\sim+} = (E^{\sim+}_1, E^{\sim+}_2, E^{\sim+}_3, \ldots., E^{\sim+}_n), \text{ where } E^{\sim+}_n = (c, c, c) \text{ and } c = \underset{i}{Max}\{c''_{ij}\},\ i = 1, 2, \ldots, m;\ j = 1, 2, \ldots, n; \tag{18}$$

$$E^{\sim} = (E^{\sim}_1, E^{\sim}_2, E^{\sim}_3, \ldots., E^{\sim}_n), \text{ where } E^{\sim}_n = (a, a, a) \text{ and } a = \underset{i}{Min}\{a''_{ij}\},\ i = 1, 2, \ldots..,m;\ j = 1, 2, \ldots, n; \tag{19}$$

**Step4: Determine the distance measures to ideal solutions**, since the $E^{\sim+}$ and $E^{\sim}$ are still TFN, we calculate $D_i^+$, the distance between the target alternative $i\ (E^{\sim}_{i,})$and the best condition in $E^{\sim+}$ from the attacker perspective as given in Eq. 20, and $D_i^-$, the distance between the target alternative $i(E^{\sim}_{i,})$ and the worst condition in $E^{\sim}$ as given in Eq. 21.

$$D_i^+ = \sum_{k=1}^{n} dist(E_k^{\sim+}, E_{i,k}^{\sim}), \qquad i = 1, 2, 3, \ldots..m \tag{20}$$

$$D_i^- = \sum_{k=1}^{n} dist(E_k^{\sim-}, E_{i,k}^{\sim}), \qquad i = 1, 2, 3, \ldots..m \tag{21}$$

Where $dist(E_k^{\sim+}, E_{i,k}^{\sim})$ and $dist(E_k^{\sim-}, E_{i,k}^{\sim})$ are calculated using the distance equation of TFN described in Section 3.2.1.

**Step5: Calculate the cost and benefits of the attacker's actions.** Based on the $D_i^+$ and $D_i^-$ distances, we compute the similarity to the worst condition as the cost of the attacker's actions ($Atc_{Cost}$) as shown in Eq. 22.

$$Atc_{Cost}(i) = \frac{D_i^-}{D_i^+ + D_i^-} \tag{22}$$

And the similarity to the best condition as the benefit of the attacker's actions ($Atc_{benefit}$) as shown in Eq. 23.

$$Atc_{benefit}(i) = \frac{D_i^+}{D_i^+ + D_i^-} \quad (23)$$

Where $i \in \{1,2\ldots,m\}$ is the actions the attacker can choose from $m$ possible actions.

## 4. RESULTS

To evaluate the new FuzVAA, we use the same 3GPP architecture that is deployed in our testbed, see Fig. 10, which we used to evaluate the original VAA with the same generated attack graph so that we can compare the two approaches to each other. This architecture is based on the concepts of control and user planes split, service base architecture, and network slicing. Their main network functionalities are the Network Slice Selection Function (NSSF), the Authentication Server Function (AUSF), the Unified Data Management (UDM), the Access and Mobility Management Function (AMF), the Session Management Function (SMF), the Policy Control Function (PCF), the Application Function (AF), the User Equipment (UE), the Radio Access Network (RAN), the User Plane Function (UPF), and the Data Network (DN). A two-level SDN controllers hierarchy bridges between the functions of the control and user planes, specifically, between the SMFs and the UPFs. The 5G core NFs are implemented as VNFs in an NFVI in which the SDN Controllers are virtualized and implemented. Fig. 11 shows the exploited assets in this case study.

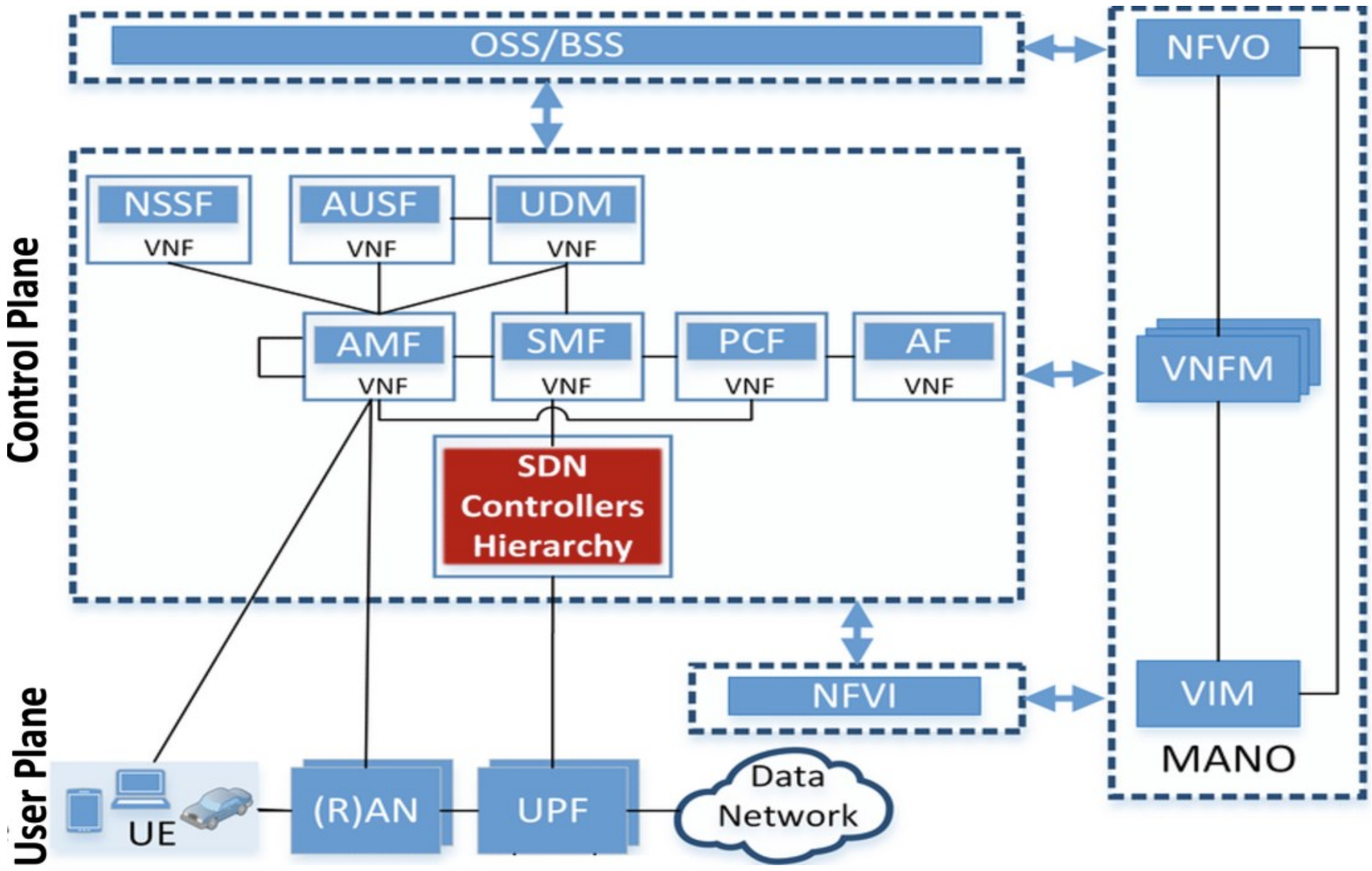

Fig. 10: The 5G Core-based 3GPP planes in our testbed.

Using the Metasploit framework [43], we run some exploits based on the 5G core attack vector. These exploits target 6 vulnerabilities in the testbed namely, the CVE-2019-15083 ( allows for an XSS injection that leads to control what software is installed on the admin workstation), CVE-2013-0375(allows for remote injection of SQL code that leads to bypassing the AUSF), CVE-2019-16026 (leads to a denial of service (DoS) condition on the AMF), CVE-2004-0415 (allows for illegitimate access to portions of kernel memory that leads to illegitimate access to the SDN), CVE-2002-0392 (allows for remote execution of DoS attack that leads to disruption for the NFVI functionalities), CVE-2004-0417 (allows for an integer overflow in the CVS Apps that leads to illegitimate access to the RAN). Fig. 12 shows the attack graph that was created using the approach described in Section 3.2.2. The main target of the attacker is to access and control the RAN module using the aforementioned vulnerabilities that belong to the three attack categories described in section 3.1.3 (i.e., *I, S, P*).

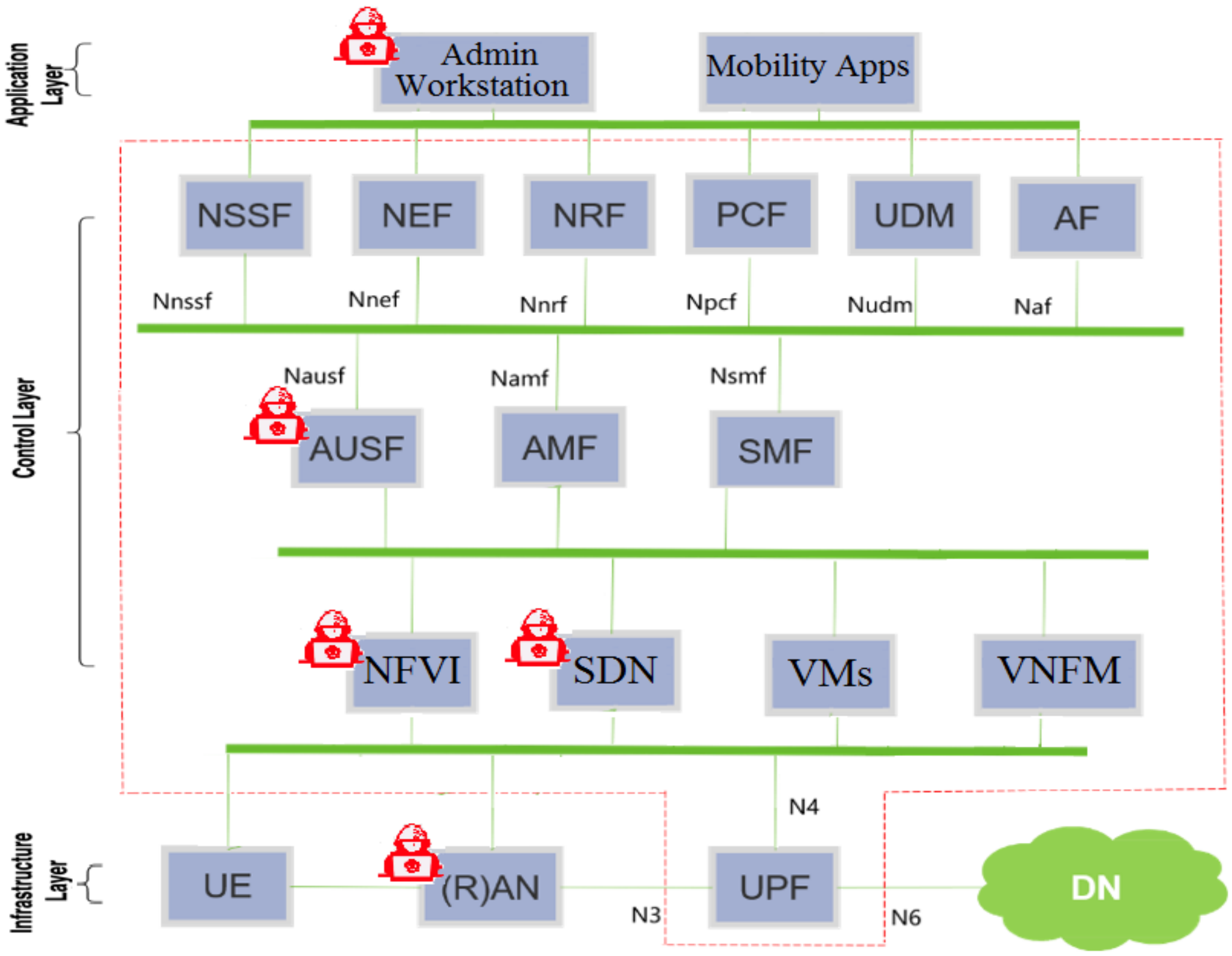

Fig.11: A case study of exploited assets in the 5G core testbed

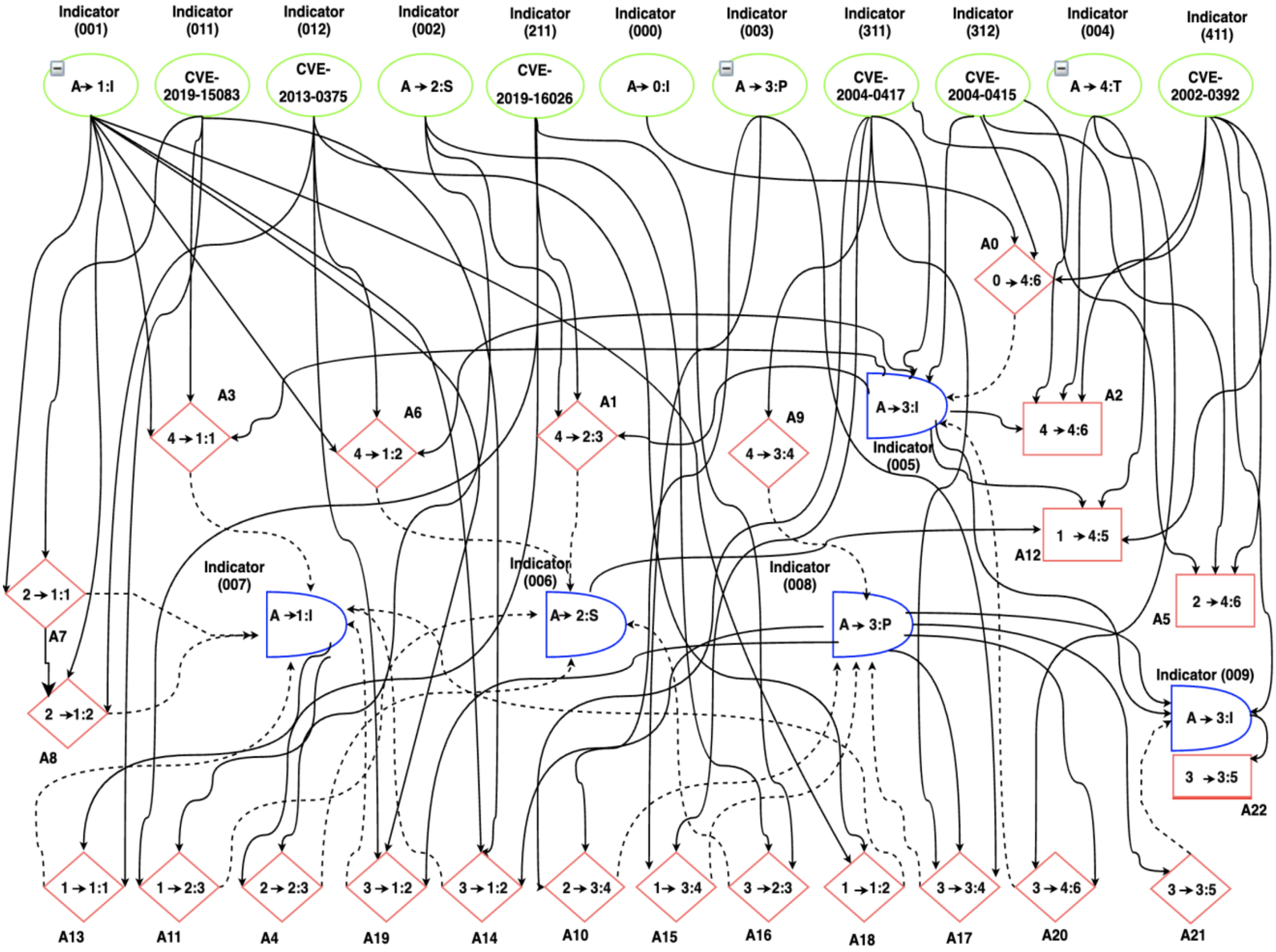

Fig.12: The attack graph with the corresponding factors' codes

Tables 3 shows an example of part of the linguistic rating variables matrix of the 5G criteria layer (for simplicity we show only 4 criterion of the CVE and the 5G dynamic factors) and the indicator layer (i.e, attacker actions, for simplicity we show only 4 actions). We set these linguistic rating variables based in the importance of the criteria and their impact on the attacker decision in the light of the information provided by the Common Vulnerability Scoring System (CVSS) described in [39] and the 5G dynamic factors described in [40].

Table 3: Linguistic Variables for the Criterion Weight Impact and the Rating

| | CVE-2004-0417 | | | | | CVE-2004-0415 | | | |
|---|---|---|---|---|---|---|---|---|---|
| | C-1 | C-2 | C-3 | C-4 | | C-1 | C-2 | C-3 | C-4 |
| **A1** | VH | H | VH | H | **A1** | AV | L | H | H |
| **A2** | H | H | AV | H | **A2** | H | H | AV | AV |
| **A3** | H | VH | VL | L | **A3** | VH | L | L | VL |
| **A4** | AV | AV | L | VL | **A4** | VH | VH | VL | H |
| | **CVE-2002-0392** | | | | | **CVE-2019-15083** | | | |
| | C-1 | C-2 | C-3 | C-4 | | C-1 | C-2 | C-3 | C-4 |
| **A1** | VH | VH | AV | AV | **A1** | H | VH | VL | AV |
| **A2** | VH | H | H | H | **A2** | AV | AV | H | H |
| **A3** | L | L | H | VH | **A3** | L | VL | H | VH |
| **A4** | AV | VH | L | VL | **A4** | AV | H | VH | VH |

Table 4 depicts the symmetric triangular fuzzy numbers that was calculated using table 2 and 3.

Table 4: The Symmetric Triangular Fuzzy Numbers

| | CVE-2004-0417 | | | | | CVE-2004-0415 | | | |
|---|---|---|---|---|---|---|---|---|---|
| | C-1 | C-2 | C-3 | C-4 | | C-1 | C-2 | C-3 | C-4 |
| **A1** | 7,9,9 | 5,7,9 | 7,9,9 | 5,7,9 | **A1** | 3,5,7 | 1,3,5 | 5,7,9 | 5,7,9 |
| **A2** | 5,7,9 | 5,7,9 | 3,5,7 | 5,7,9 | **A2** | 5,7,9 | 5,7,9 | 3,5,7 | 3,5,7 |
| **A3** | 5,7,9 | 7,9,9 | 1,1,3 | 1,3,5 | **A3** | 7,9,9 | 1,3,5 | 1,3,5 | 1,1,3 |
| **A4** | 3,5,7 | 3,5,7 | 1,3,5 | 1,1,3 | **A4** | 7,9,9 | 7,9,9 | 1,1,3 | 5,7,9 |
| | **CVE-2002-0392** | | | | | **CVE-2019-15083** | | | |
| | C-1 | C-2 | C-3 | C-4 | | C-1 | C-2 | C-3 | C-4 |
| **A1** | 7,9,9 | 7,9,9 | 3,5,7 | 3,5,7 | **A1** | 5,7,9 | 7,9,9 | 1,1,3 | 3,5,7 |
| **A2** | 7,9,9 | 5,7,9 | 5,7,9 | 5,7,9 | **A2** | 3,5,7 | 3,5,7 | 5,7,9 | 5,7,9 |
| **A3** | 1,3,5 | 1,3,5 | 5,7,9 | 7,9,9 | **A3** | 1,3,5 | 1,1,3 | 5,7,9 | 7,9,9 |
| **A4** | 3,5,7 | 7,9,9 | 1,3,5 | 1,1,3 | **A4** | 3,5,7 | 5,7,9 | 7,9,9 | 7,9,9 |

Table 5: An example of a combined decision matrix

| | C-1 | C-2 | C-3 | C-4 | | C-1 | C-2 | C-3 | C-4 |
|---|---|---|---|---|---|---|---|---|---|
| **A1** | $\tilde{x}_{11}$ | $\tilde{x}_{12}$ | $\tilde{x}_{13}$ | $\tilde{x}_{14}$ | **A1** | 3, 7.5, 9 | 1, 7, 9 | 1, 5.5, 9 | 3, 6, 9 |
| **A2** | $\tilde{x}_{21}$ | $\tilde{x}_{22}$ | $\tilde{x}_{23}$ | $\tilde{x}_{24}$ | **A2** | 3, 7, 9 | 3, 6.5, 9 | 3, 6, 9 | 3, 6.5, 9 |
| **A3** | $\tilde{x}_{31}$ | $\tilde{x}_{32}$ | $\tilde{x}_{33}$ | $\tilde{x}_{34}$ | **A3** | 1, 5.5, 9 | 1, 4, 9 | 1, 4.5, 9 | 1, 5.5, 9 |
| **A4** | $\tilde{x}_{41}$ | $\tilde{x}_{42}$ | $\tilde{x}_{43}$ | $\tilde{x}_{44}$ | **A4** | 3, 6, 9 | 3, 7.5, 9 | 1, 4, 9 | 1, 4, 9 |

Using combined matrices and Equation 22, we compute the $Atc_{Cost}$ for each possible attacker actions, see table 6. We then choose the lowest attacker efforts in the three attacking schemes (i.e., I, S, P).

Table 6: An example the computed attack cost of the attacker actions.

| Criteria/ Attribute | $D_i^+$ | $D_i^-$ | $\frac{D_i^-}{D_i^- + D_i^+}$ | $Atc_{Cost}(i)$ |
|---|---|---|---|---|
| A1 | 4.326 | 4.194 | [4.194/(4.194 + 4.326)] | 0.492 |
| A2 | 4.010 | 3.785 | [3.785/(3.785 + 4.010)] | 0.486 |
| A3 | 3.420 | 3.726 | [3.726/(3.726 + 3.420)] | 0.521 |
| A4 | 1.333 | 7.485 | [7.485/(7.485 + 1.333)] | 0.849 |

### 4.1 Compare the accuracy of the VAA with the Nessus.

The underlying idea behind the VEA-bility metric [41] is that the security of a network is influenced by many factors, including the severity of existing vulnerabilities, distribution of services, connectivity of hosts, and possible attack paths. These factors are modeled into three network dimensions: Vulnerability, Exploitability, and Attackability. The overall VEA-bility score, a numeric value in the range [0,10], is a function of these three dimensions, where a lower value implies better security. The VEA-bility metric uses data from three sources: the 5G core testbed topology, attack graphs, and scores as assigned by the Common Vulnerability Scoring System (CVSS) [40]. To adjust the VEA-bility metric to validate the accuracy of the vulnerability assessment of the VAA and Nessus [42], we modify this metric by replacing the asset Attackability factor with the $Atc_{Cost}(i)$ value at Eq. 22 for each set of actions *i*. We let each vulnerability *v,* that corresponds to a set of actions *i*, have an impact score, exploitability score, and temporal score as defined by the CVSS. An impact and exploitability subscores are automatically generated for each common vulnerabilities identifier based on its CVE name defined by the CVSS, whereas the temporal score requires user input. We then define the severity, *S*, of a vulnerability to be the average of the impact and temporal scores, Eq. (24):

$$S(v) = (Impact\ Score(v) + Temporal\ Score(v)) / 2 \quad (24)$$

The Vulnerability score (*V*) of a 5G core testbed asset, e.g., UE, MEC server, SDN, NFV, etc is an exponential average of the severity scores of the vulnerabilities on the 5G core asset, or 10, whichever is lower. The asset Exploitability score (*E*) is the exponential average of the exploitability score for all asset vulnerabilities multiplied by the ratio of network services on the asset. The asset Attackability score (*A*) refers to the toral CP values for all vulnerabilities at a certain asset. The Attackability score is multiplied by a factor of 10 to produce a number in the range [0,10], ensuring that all dimensions have the same range. For an asset, *a*, let *v* be an asset vulnerability. We then define the three asset dimensions as shown in Eq. 25, Eq. 26. and Eq.27:

$$V(a) = min(10, ln \sum e^{S(v)}) \quad (25)$$

$$E(a) = (min(10, ln \sum e^{Exploitability\ Score(v)})) (\#\ services\ on\ a)/(\#\ network\ services) \quad (26)$$

$$A(t) = (10) * \sum_{i=1}^{n} a_{CP(e_i)} \quad (27)$$

The overall VEA-bility equation for an asset *a is* then computed as in Eq. (28).

$$VEA\text{-}bility_a = 10 - ((V+E+A)_a / 3) \quad (28)$$

### 4.2 Compare the accuracy of the FuzVAA with the original VAA and the Nessus.

To compare the accuracy of the FuzVAA, original VAA and Nessus using the proposed VEA-bility metric, we developed an extensive set of attacks scenarios and used the vulnerabilities observed by the Nessus scan [42], original VAA, and FuzVAA and results after running the attacks scenarios. Fig. 13 shows the overall average VEA-bility scores observed in our experiments for the 5G core testbed assets. A higher score indicates a more secure configuration, which we call more "VEA-ble". Fig. 13 shows that the FuzVAA, on average, is 5.37% and 34.97% more VEA-ble than the original VAA and Nessus respectively.

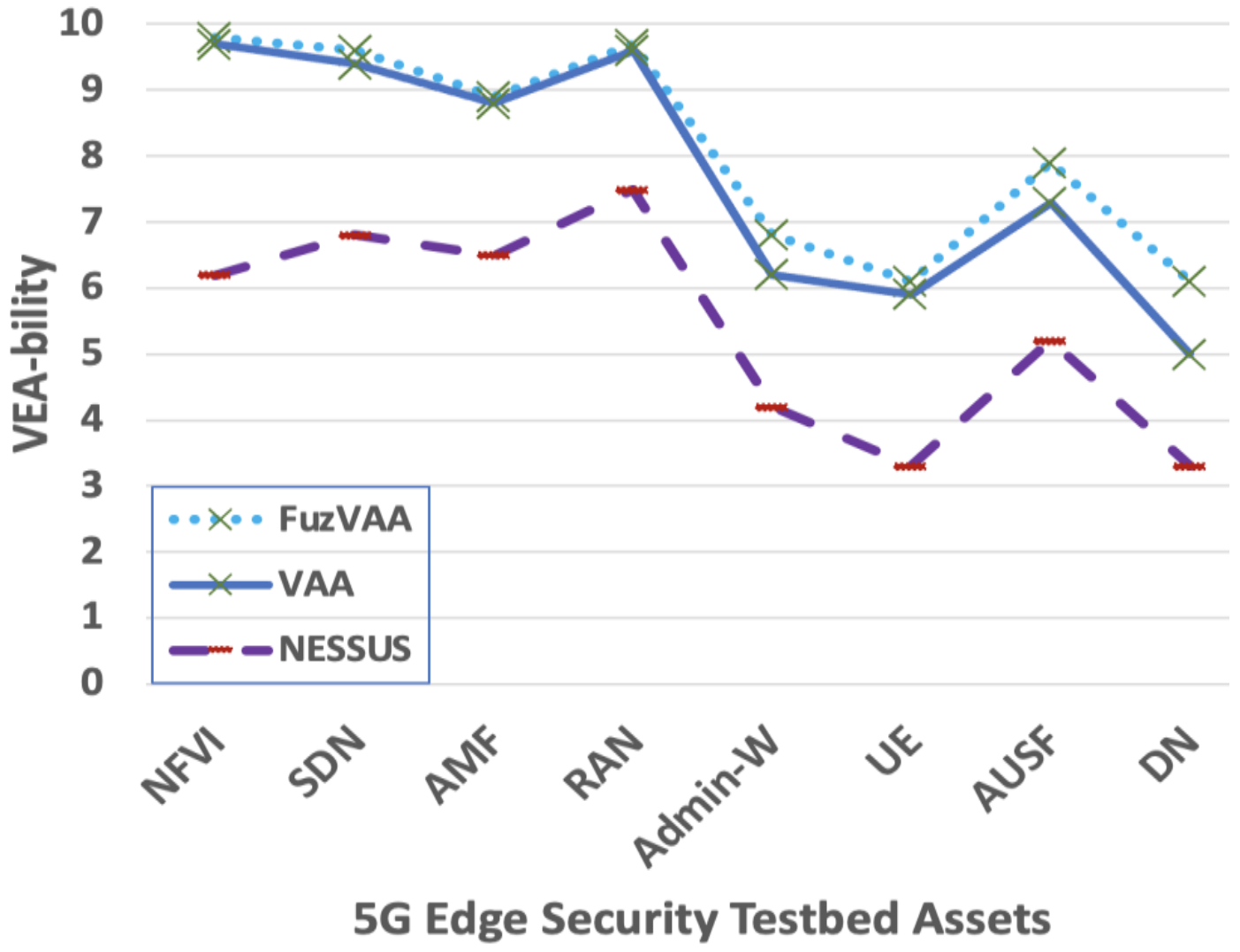

Fig. 13: The VEA-bility metric of the FuzVAA, original VAA, and Nessus.

To compare the performance of the FuzVAA, original VAA [2], and Nessus, we run the experiment based on the above-mentioned 6 vulnerabilities. Fig. 14 shows the performance of the FuzVAA, original VAA, and Nessus in milliseconds. The original VAA, on average, outperforms Nessus and FuzVAA by 27.14% and 6.58%, respectively. The original VAA takes 6151ms to compute the attack cost related to all possible paths of the 6 vulnerabilities while Nessus and the original VAA take 8445ms and 6556ms, respectively, to assess the same 6 vulnerabilities. The FuzVAA outperforms Nessus by 22.36%. This shows that our VAA introduces a more scalable and faster assessment.

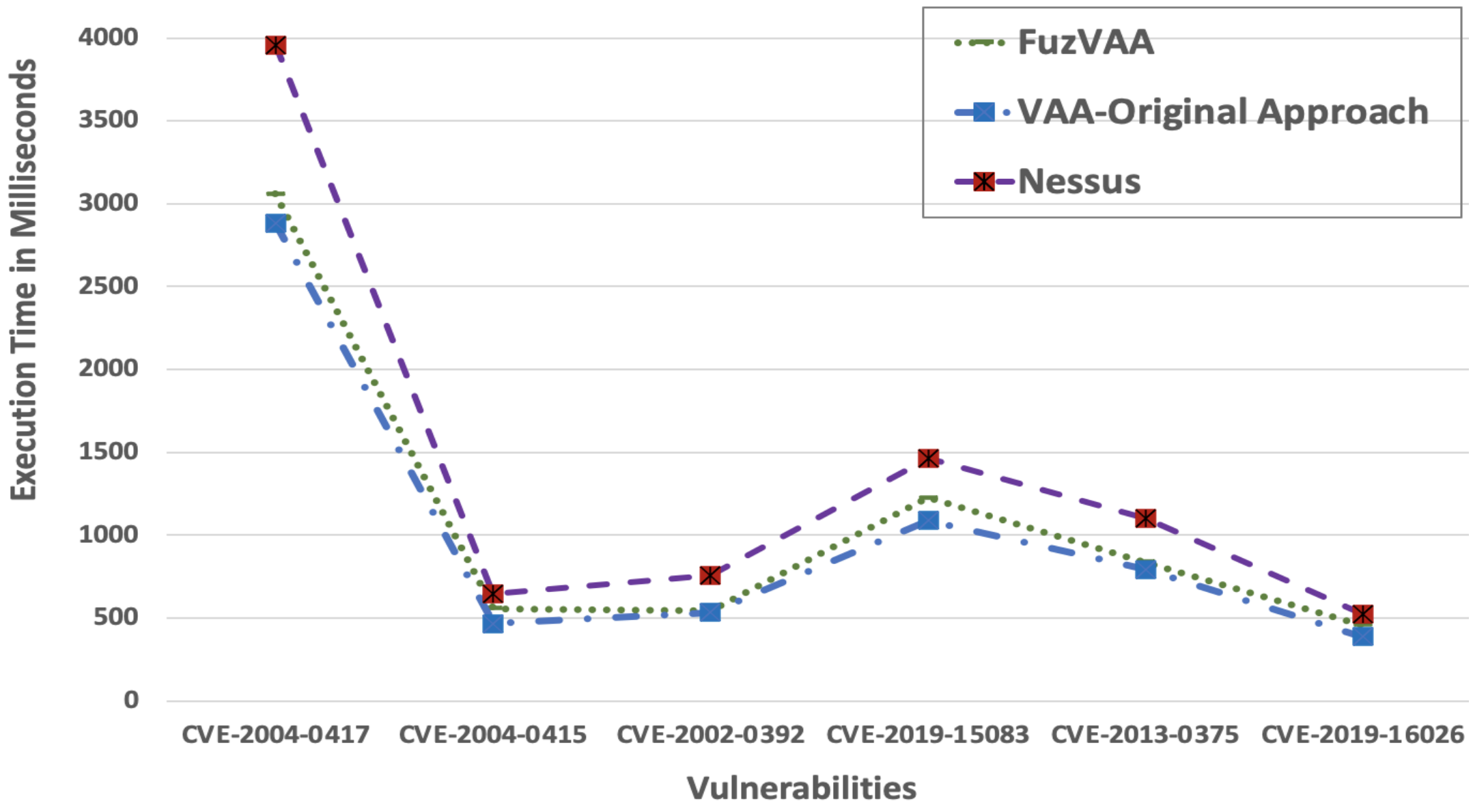

Fig. 14: The performance metric of the FuzVAA, original VAA, and Nessus,

## 5. CONCLUSION

The 5G system improves the bandwidth and capabilities of the current telecommunication infrastructure. However, it introduces new threats and attacks. In this paper, we introduced the FuzVAA , a scalable and accurate vulnerability analysis approach that was tested and evaluated using our new developed 5G core testbed. The experiment results depict that FuzVAA successfully analyzed the vulnerabilities with a low error rate. Experiment results show that, on average, it 5.37% and 34.97 more VEA-ble than the original VAA and Nessus respectively. The original VAA, on average, outperforms Nessus and FuzVAA by 27.14% and 6.58%, respectively. We also introduced our security deployment model which orchestrates the security services using network slices. This model connects all IDS agents through the radios using FlexRAN to the central security management server to analyze and make decisions about the severity of the IDS alerts. We run a case study to validate the security testbed. The experiment showed that the deployment model works very well and can be used to conduct various security experiments.

In future work, we will integrate the FuzVAA with an autonomous intrusion response system that considers the vulnerability assessment values of FuzVAA to deploy countermeasures against cyber-attacks. Furthermore, we will compare the accuracy and performance of the FuzVAA against our previously developed VAA based Hexagonal fuzzy method [50].

## REFERENCES

[1] "The DoD's 5G strategy implementation plan", https://apps.dtic.mil/sti/pdfs/AD1118833.pdf

[2] Kholidy, H.A.; Karam, A.; Sidoran, J.L.; Rahman, M.A. 5G Core Security in Edge Networks: A Vulnerability Assessment Approach. In Proceedings of the 26th IEEE Symposium on Computers and Communications (ISCC), Athens, Greece, 5–8 September 2021; pp. 1–6. [Google Scholar] [CrossRef]

[3] Behzadian, M., Khanmohammadi Otaghsara, S., Yazdani, M., & Ignatius, J. (2012). A state-of-the-art survey of TOPSIS applications. Expert Systems with Applications, 39(17), 13051–13069. doi:10.1016/j.eswa.2012.05.056

[4] Abhishek K., Bikash Sah, Arvind R. Singh, Yan Deng, Xiangning He, Praveen Kumar, Ramesh Bansal, "Chapter 1 - Multicriteria decision-making methodologies and their applications in sustainable energy system/microgrids", Decision Making Applications in Modern Power Systems, 2020, Pages 1-40, ISBN 9780128164457.

[5] OpenStack, https://www.openstack.org/

[6] Open-Source MANO (OSM), https://osm.etsi.org/

[7] Open5GS, https://open5gs.org/

[8] FlexRAN (Mosaic5G), https://mosaic5g.io/flexran/

[9] OSSIM, https://cybersecurity.att.com/products/ossim

[10] SNORT, https://www.snort.org/

[11] OSSEC. https://www.ossec.net/

[12] "Open Networking Foundation", https://opennetworking.org/

[13] "ETSI NFV ARCHITECTURE & INTERFACES", https://docbox.etsi.org/ISG/NFV/Open/other/Tutorials/2017-Tutorials-NFV_World_Congress_San_Jose/RX11449_ETSI_NFV_IFA_presentation_at_NFV_World_Congress_2017.pdf

[14] "OSM Kubernetes cluster from an OSM Network Service", https://osm.etsi.org/docs/user-guide/15-k8s-installation.html

[15] I. H. Abdulqadder, D. Zou, I. T. Aziz, B. Yuan, W. Li, "Secsdn-cloud: Defeating vulnerable attacks through secure software-defined networks", IEEE Access (2018).

[16] D. Yin, L. Zhang, K. Yang, A ddos attack detection and mitigation with software-defined internet of things framework, IEEE Access (2018).

[17] Z. Fan, Y. Xiao, A. Nayak, C. Tan, An improved network security situation assessment approach in software defined networks, Peer-to-Peer Networking and Applications 12 (2) (2019).

[18] V. Varadharajan, K. Karmakar, U. Tupakula, M. Hitchens, A policy-based security architecture for software-defined networks, IEEE Transactions on Information Forensics and Security 14 (4) (2019).

[19] H. Li, F. Wei, H. Hu, Enabling dynamic network access control with anomaly-based ids and sdn, in: Proceedings of the ACM International Workshop on Security in Software Defined Networks & Network Function Virtualization, ACM, 2019, pp. 13–16.

[20] J. Yao, Z. Han, M. Sohail, L. Wang, A robust security architecture for sdn-based 5g networks, Future Internet 11 (4) (2019) 85. 941.
[21] ISO/IEC 27002: Code of Practice for Information Security Management. 2005. http://www.iso.org/iso/catalogue_detail?csnumber=54533
[22] National Institute of Standards and Technology. NIST-SP800 Series Special Publications on Computer Security.
[23] I. Abdulqadder, D. Zou, I. Aziz, B. Yuan, W. Dai, Deployment of robust security scheme in sdn based 5g network over nfv enabled cloud environment, 943 IEEE Transactions on Emerging Topics in Computing.
[24] Kholidy, H. A. (2020), "Autonomous mitigation of cyber risks in the Cyber–Physical Systems", doi:10.1016/j.future.2020.09.002, Future Generation Computer Systems,Volume 115, 2021, Pages 171-187, ISSN 0167-739X, https://doi.org/10.1016/j.future.2020.09.002.
[25] Kholidy, H.A., Baiardi, F., Hariri, S., et al.: "A hierarchical cloud intrusion detection system: design and evaluation", Int. J. Cloud Comput., Serv. Archit. (IJCCSA), 2012, 2, pp. 1–24.
[26] Kholidy, H.A., "Autonomous mitigation of cyber risks in the Cyber–Physical Systems", Future Generation Computer Systems, Volume 115, 2021, Pages 171-187, ISSN 0167-739X, https://doi.org/10.1016/j.future.2020.09.002.
[27] Kholidy, H.A., Abdelkarim Erradi, Sherif Abdelwahed, Fabrizio Baiardi, "A risk mitigation approach for autonomous cloud intrusion response system", in Journal of Computing, Springer, DOI: 10.1007/s00607- 016-0495-8, June 2016.
[28] Kholidy, H.A., Abdelkarim Erradi, "A Cost-Aware Model for Risk Mitigation in Cloud Computing SystemsSuccessful accepted in 12th ACS/IEEE International Conference on Computer Systems and Applications (AICCSA), Marrakech, Morocco, November, 2015.
[29] Kholidy, H.A., Ali Tekeoglu, Stefano Iannucci, Shamik Sengupta, Qian Chen, Sherif Abdelwahed, John Hamilton, "Attacks Detection in SCADA Systems Using an Improved Non-Nested Generalized Exemplars Algorithm", the 12th IEEE International Conference on Computer Engineering and Systems (ICCES 2017), December 19-20, 2017.
[30] Kholidy, H.A., Fabrizio Baiardi, "CIDS: A framework for Intrusion Detection in Cloud Systems", in the 9th Int. Conf. on Information Technology: New Generations ITNG 2012, April 16-18, Las Vegas, Nevada, USA. http://www.di.unipi.it/~hkholidy/projects/cids/
[31] Hisham A. Kholidy, Abdelkarim Erradi, Sherif Abdelwahed, Abdulrahman Azab, "A Finite State Hidden Markov Model for Predicting Multistage Attacks in Cloud Systems", in the 12th IEEE International Conference on Dependable, Autonomic and Secure Computing (DASC), Dalian, China, August 2014.
[32] Kholidy, H. A., & Erradi, A. (2019). VHDRA: A Vertical and Horizontal Intelligent Dataset Reduction Approach for Cyber-Physical Power Aware Intrusion Detection Systems. Security and Communication Networks, 2019, 1–15. doi:10.1155/2019/6816943
[33] Kholidy, H.A., "Detecting impersonation attacks in cloud computing environments using a centric user profiling approach", Future Generation Computer Systems, Volume 115, issue 17, December 13, 2020, Pages 171-187, ISSN 0167-739X, https://doi.org/10.1016/j.future.2020.12.
[34] Hisham A. Kholidy, "Detecting impersonation attacks in cloud computing environments using a centric user profiling approach", Future Generation Computer Systems, Volume 117, 2021, Pages 299-320, ISSN 0167-739X, https://doi.org/10.1016/j.future.2020.12.009.
[35] Kholidy, Hisham A.: 'Correlation-based sequence alignment models for detecting masquerades in cloud computing', IET Information Security, 2020, 14, (1), p.39-50, DOI: 10.1049/iet-ifs.2019.0409.
[36] H. A. Kholidy, "Towards A Scalable Symmetric Key Cryptographic Scheme: Performance Evaluation and Security Analysis," 2019 2nd International Conference on Computer Applications & Information Security (ICCAIS), Riyadh, Saudi Arabia, 2019, pp. 1-6, doi: 10.1109/CAIS.2019.8769494.
[37] Kholidy, H.A., Fabrizio Baiardi, Salim Hariri: 'DDSGA: A Data-Driven Semi-Global Alignment Approach for Detecting Masquerade Attacks'. The IEEE Transaction on Dependable and Secure Computing, 10.1109/TDSC.2014.2327966, pp:164–178, June 2015.
[38] Kholidy, H.A., Fabrizio Baiardi, "CIDD: A Cloud Intrusion Detection Dataset For Cloud Computing and Masquerade Attacks ", in the 9th Int. Conf. on Information Technology: New Generations ITNG 2012, April 16-18, Las Vegas, Nevada, USA. http://www.di.unipi.it/~hkholidy/projects/cidd/
[39] Bräuning, F., & Koopman, S. J. (2019). The dynamic factor network model with an application to international trade. Journal of Econometrics. doi:10.1016/j.jeconom.2019.10.007
[40] https://www.first.org/cvss/specification-document
[41] Tupper M, Zincir-Heywood A (2008) VEA-bility security metric: a network security analysis tool. In: Proc IEEE Third Int'l Conf. Availability, Reliability and Security.
[42] Nessus Vulnerability Scanner: http://www.nessus.org.
[43] The Metasploit framework, https://www.metasploit.com/
[44] "Ksniff: packet capture at pod level for k8s", https://kubesandclouds.com/index.php/2021/01/20/ksniff/
[45] The LOIC tool, https://www.imperva.com/learn/ddos/low-orbit-ion-cannon/
[46] The Minikube , https://minikube.sigs.k8s.io/docs/start/\

[47] Wei Huang, Gazi Shakhawat Hossain, Ayrin Sultana, Md. Rasel, Md. Shahinur Rahman, "Applying Fuzzy Technique for Order Preference by Similarity to Ideal Solution (TOPSIS) in the Selection of Best Candidate: A Case Study on Interview Performance", British Journal of Economics, Finance and Management Sciences, February 2020, Vol. 17 (1).
[48] Kholidy, H.A. Multi-Layer Attack Graph Analysis in the 5G Edge Network Using a Dynamic Hexagonal Fuzzy Method. Sensors 2022, 22, 9. https://doi.org/10.3390/s22010009
[49] Haque, N.; Rahman, M.; Chen, D.; Kholidy, H. BIoTA: Control-Aware Attack Analytics for Building Internet of Things. In Proceedings of the 18th IEEE International Conference on Sensing, Communication and Networking (SECON), Rome, Italy, 6–9 July 2021.
[50] Kholidy, H.A. Multi-Layer Attack Graph Analysis in the 5G Edge Network Using a Dynamic Hexagonal Fuzzy Method. Sensors 2022, 22, 9. https://doi.org/10.3390/s22010009